\documentclass[letter,11pt]{article}
\usepackage{graphicx}
\usepackage{amssymb}
\usepackage{epstopdf}
\usepackage{url}
\usepackage{anysize}
\marginsize{1in}{1.in}{1.25in}{1.25in}

\newcommand{\be}{\begin{equation}}
\newcommand{\bea}{\begin{eqnarray}}
\newcommand{\beq}[1]{\begin{equation}\label{#1}}

\newcommand{\ee}{\end{equation}}
\newcommand{\eea}{\end{eqnarray}}
\newcommand{\eeq}{\end{equation}}
\newcommand{\lsim}{\!\mathrel{\hbox{\rlap{\lower.55ex \hbox{$\sim$}} \kern-.34em \raise.4ex \hbox{$<$}}}}
\newcommand{\gsim}{\!\mathrel{\hbox{\rlap{\lower.55ex \hbox{$\sim$}} \kern-.34em \raise.4ex \hbox{$>$}}}}

\begin{document}
 
\begin{titlepage}
\flushright{MCTP-08-06\\}
\vspace{1in}
\begin{center}
{\Large \bf Testing a $U(1)$ Solution to the $\mu$ Problem}

\vspace{0.5in}
{\bf Timothy Cohen and Aaron Pierce}

\vspace{0.2cm}
{\it Michigan Center for Theoretical Physics (MCTP)\\Randall Laboratory, Physics Department, University of Michigan, \\ Ann Arbor, MI 48109}

\end{center}
\vspace{0.8cm}
\begin{abstract}
We discuss the collider phenomenology of TeV $Z^{\prime}$ gauge bosons related to the absence of a bare $\mu$-term in the superpotential.  Decays of the type  $Z^{\prime} \rightarrow$ Higgsinos can directly test whether a gauge symmetry is responsible for forbidding the Higgsino mass.  Decays to multi-lepton final states may allow these signatures to be observed at the Large Hadron Collider.  We comment on whether it will be possible to  state definitively that the $\mu$-term is forbidden via a gauge symmetry.
\end{abstract}
\end{titlepage}

\section{Introduction}
Supersymmetry (SUSY) remains one of the most attractive possibilities for new physics at the weak scale.  It addresses the unnatural Higgs boson mass, provides a viable Dark Matter candidate, and leads to the apparent unification of couplings at an energy scale not too far from the Planck scale.

In its simplest incarnation, the Minimal Supersymmetric Standard Model (MSSM)  contains a puzzle. The superpotential contains a dimensionful parameter, $\mu$:
\begin{equation}
\label{eqn:mudef}
W= \mu H_{u} H_{d}.
\end{equation}
To achieve natural electroweak symmetry breaking, this parameter must be of order the weak scale, which is set by the scale of the SUSY breaking soft masses.  There is no a priori reason to expect a relationship between supersymmetric and SUSY breaking parameters.  This is the ``mu problem'' of the MSSM, see \cite{Polonsky} for a review.

One attractive solution is to forbid the $\mu$ parameter of Eqn.~(\ref{eqn:mudef}), and to generate an effective $\mu$-term dynamically.   The trick is to then arrange for the dynamics (presumably connected to SUSY breaking) to produce a $\mu$-term of the right size.  One approach is to add operators suppressed by a high scale, i.e. the Giudice-Masiero mechanism \cite{Giudice:1988yz}.  A second is to modify the low energy effective theory, adding an additional light Standard Model (SM) singlet state, $S$.  This is the approach of the Next-to-Minimal Supersymmetry Standard Model  (NMSSM)\cite{NMSSM}, where the $\mu$-term is generated via the superpotential term:
\begin{equation}
\label{eqn:mueff}
W = \lambda S H_{u} H_{d}.
\end{equation}
When $S$ acquires a vacuum expectation value (vev), this yields $\mu_{eff} = \lambda \langle S \rangle$. But what is the symmetry that forbids the bare $\mu$-term of Eqn.~(\ref{eqn:mudef})?  One can appeal to a global symmetry  (e.g.,  a ${\bf Z_{3}})$, but such symmetries often lead to cosmological difficulties, whose solutions tend to destabilize the hierarchy\cite{DomainWalls}.  Another possibility is to charge $H_{u}$, $H_{d}$ and $S$ under a new $U(1)^{\prime}$ gauge symmetry.  By taking the charges of the Higgs supermultiplets such that 
\begin{eqnarray}
Q_{H_{u}}^{\prime} + Q_{H_{d}}^{\prime} &\neq& 0, \nonumber\\  
Q_{H_{u}}^{\prime} + Q_{H_{d}}^{\prime} + Q_{S}^{\prime} &=& 0,
\label{eqn:forbidcharges}
\end{eqnarray} 
gauge invariance can simultaneously forbid the bare $\mu$-term of Eqn.~(\ref{eqn:mudef}) while allowing the desired term of Eqn.~(\ref{eqn:mueff}).  This simple observation is perhaps one of the best motivations for building models with a $U(1)^{\prime}$ gauge boson at the weak scale.

Suppose that the Large Hadron Collider (LHC) discovers a $Z^{\prime}$ gauge boson, presumably through the process $ p\,\bar{p} \rightarrow \ell^{+} \ell^{-}$. Such a discovery would lead to a reprise of I.I. Rabi's famous comment regarding the muon: ``Who ordered that?''.  For while $Z^{\prime}$ gauge bosons are motivated by many extensions to the standard model, e.g. Grand Unified Theories (GUTs) and string constructions (see \cite{Langacker2008} for references and a recent review of these and other motivations), it is difficult to motivate why the $Z^{\prime}$ should appear at the TeV scale.  If, however, the gauge symmetry associated with the $U(1)^{\prime}$ is what forbids the $\mu$-term, then the coincidence of the weak scale and the $Z^{\prime}$ mass is explained.  Because $\mu$-term must be at the electroweak scale to explain natural electroweak symmetry breaking, the $\langle S \rangle$ is of order the weak scale.  Assuming this vev dominates the $Z^{\prime}$ mass,  the $U(1)^{\prime}$ is unbroken down to this scale as well.  In this case, the $\mu$-term ordered the new $Z^{\prime}$.

Many studies discussing the measurement of $Z^{\prime}$ properties exist in the literature (see \cite{Langacker2008, RizzoTASI} for reviews).  Most focus on the detailed examination of leptonic final states or rare decays to gauge bosons.  In this paper, we discuss the possibility of observing the direct decay of a TeV scale $Z^{\prime}$ to Higgsinos.  Such decays indicate that the Higgsinos (and hence the Higgs supermultiplets) are charged under the new gauge symmetry as necessitated by Eqn.~(\ref{eqn:forbidcharges}).  These decays represent a smoking gun, perhaps the most direct way to show that the $U(1)^{\prime}$ symmetry is related to forbidding the $\mu$-term.

In the next section, we briefly review some model building considerations related to attempting to forbid the $\mu$-term via a gauge symmetry.  We also introduce the benchmark $U(1)^{\prime}$ model that will be used in the collider studies that follow.  In section \ref{sec:Collider}, we discuss the possibility of observing the decays $Z^{\prime} \rightarrow  \tilde{\chi}_{i}^{0} \tilde{\chi}_{j}^{0}$ at the LHC.  We rely heavily on decays of the type  $\tilde{\chi}_{j}^{0} \rightarrow \tilde{\chi}_{1}^0 \ell^{+} \ell^{-}$, a nearly background free channel. In section \ref{sec:Implications}, we discuss what can be learned from studying these decays, and whether we will be able to say definitively that the $\mu$-term is forbidden by the new gauge symmetry.  Finally, we conclude.

\section{Forbidding the $\mu$-term with a $Z^{\prime}$}
While adding a $U(1)^{\prime}$ symmetry  to the MSSM is a well-motivated method for initially forbidding and subsequently generating the $\mu$-term dynamically, it does introduce many model-building difficulties.  Charging the Higgs bosons under the $U(1)^{\prime}$ while simultaneously allowing Yukawa couplings, forces the SM matter fields to also be charged under the $U(1)^{\prime}$.  This induces new conditions to avoid mixed-anomalies between the $U(1)^{\prime}$ and the SM symmetry groups.  It is challenging to satisfy these conditions while simultaneously maintaining gauge coupling unification and avoiding introducing new ``$\mu$-like" terms for exotic matter \cite{Wells, Arvanitaki}.  We briefly discuss this tension, along with other model building challenges before settling on a choice for the charges of our $U(1)^{\prime}$.

To avoid disrupting gauge coupling unification one can restrict new particles to come in complete GUT multiplets or to be singlets under the SM. Canceling mixed anomalies indicates the presence of particles with non-trivial $SU(3)$ and $SU(2)_{L}$ charges.  For example, the $U(1)^{\prime}\mathrm{-}SU(3)^2$ anomaly requires a pair of exotic quarks ($D'$) while the $U(1)^{\prime}\mathrm{-}SU(2)_{L}^2$ anomaly requires a pair of exotic leptons ($L'$).   One economical way to implement this new matter is by introducing new $\textbf{5}+\overline{\textbf{5}}$ representations where the $D'$s can have different $U(1)^{\prime}$ charges from the $L'$s.  However, having $D'$ and $L'$ fields with different $U(1)^{\prime}$ charges calls the simplest GUT interpretations into question.

If one chooses to introduce only one set of $\textbf{5}+{\bf \overline{5}}$s, the four $U(1)^{\prime}\mathrm{-}\mathrm{SM}$ anomaly conditions fix their $U(1)^{\prime}$ charges.  These new fields require a mass, via the vacuum expectation value of a new singlet(s).   In turn, these singlets need a mass.    Also, the $U(1)^{\prime \, 3}$ and the $U(1)^{\prime}$-gravitational anomalies must be cancelled.  Almost without fail, this leads to additional SM singlets (often with irrational $U(1)^{\prime}$ charges) \cite{Wells}.   The model building can rapidly become baroque.\footnote{The new and now fairly complex scalar potential must be addressed as well, since there will almost certainly be D-Flat directions to worry about.  Finally, the $D'$s and $L'$s must decay fast enough to satisfy cosmological constraints without introducing baryon number violation that would lead to a too--short proton lifetime. See, e.g., \cite{Arvanitaki} for some discussions of these points within the setting of a particular model.}

The full implementation of the singlet/exotic sector can affect the collider phenomenology.  After all, the singlet vevs all contribute to the mass of the $Z^{\prime}$.   If ``too many'' SM singlets get large vevs, the $Z^{\prime}$ can be pushed to a mass that makes detailed observations difficult.  Also, these SM singlet--superfields, if light, can modify the neutralino sector.   So, while there might well be interesting phenomenology associated with the implementation of a particular $Z^{\prime}$ model, we choose instead to consider a decoupling limit of sorts where the singlets do not affect the details of the neutralino sector, nor are they present in the decays of the $Z^{\prime}$.  Furthermore, we assume all colored/charged exotics are sufficiently heavy, so that they are not produced in $Z^{\prime}$ decays.  It would be interesting to relax these assumptions.  If light, the new exotics will present exciting phenomenological opportunities \cite{Kang:2007ib}, including the possibility of long lived heavy colored particles, reminiscent of Split Supersymmetry or Hidden Valley models.   One could potentially investigate $Z^{\prime}$ decays directly to these states.  We leave a detailed study of this possibility to future work.

Here we recognize the challenges of embedding a $U(1)^{\prime}$ symmetry in a consistent model, but will choose to be agnostic about the specifics of how these problems are solved.  To readily achieve our decoupling limit, we will follow an approach loosely motivated by $E_{6}$ GUTs, taken, e.g. in the recent work of \cite{King}.  For this model the charges of NMSSM fields under the  $U(1)^{\prime}$ are given in Table \ref{tab:ZprimeCH}.   Changing the charges of the fields under the gauge symmetry will affect the details of the phenomenology we discuss here, but will not affect the gross features -- nor the basic fact that one should look for decays of the type $Z^{\prime} \rightarrow$ Higgsinos.

\begin{table}[h]
\begin{center}
\begin{tabular}{|l|l|l|l|l|l|l|l|l|}
\hline
Matter  & $Q$ & $U$ & $D$ & $L$ & $E$ & $H_u$ & $H_d$& $S$\\
\hline
$(2\sqrt{10}) \times Q'$ & 1 & 1 & 2 & 2 & 1 & -2 & -3 & 5 \\
\hline
\end{tabular}
\end{center}
\caption{The benchmark $Z^{\prime}$ charges used throughout.  This choice corresponds to the $E_6$ charges where the right handed neutrino is neutral under the $U(1)^{\prime}$.}
\label{tab:ZprimeCH}
\end{table}

\subsection{How light can the $Z^{\prime}$ be?}
The signal that we will discuss in the following section will be statistics limited.  Thus, it will be most visible for a light $Z^{\prime}$.  However, even neglecting direct searches, the $Z^{\prime}$ cannot be arbitrarily light due to precision electroweak constraints. After all, in the models that we consider here, the Higgs fields are necessarily charged under the $U(1)^{\prime}$ and thereby introduce $Z^0-Z^{\prime}$ mixing.

To calculate the $Z-Z^{\prime}$ mixing, one must also account for possible kinetic mixing between the hypercharge gauge boson and the $U(1)^{\prime}$ gauge boson, leading to a covariant
derivative of the form \cite{HaberChoi}
\begin{eqnarray}
D_{\mu}&=&\partial_{\mu}+i\,g_Y\,Y\,B_{\mu}+i\,g'(\frac{1}{\cos{\chi}}Q'-\frac{g_Y}{g'}\,\tan{\chi}\,Y)B_{\mu}'\\
      &=&\partial_{\mu}+i\,g_Y\,Y\,B_{\mu}+i\,g'\,Q^{\prime\,\,\mathrm{mixed}}\,B_{\mu}'
\end{eqnarray}
where $g_Y$ is the hypercharge coupling constant, $g'$ is the $U(1)^{\prime}$ coupling
constant, $Y$ is the hypercharge generator, $Q'$ is the $U(1)^{\prime}$ generator,
$Q^{\prime\,\,\mathrm{mixed}}$ is the resulting generator due to kinetic mixing,
$B_{\mu}$ is the hypercharge gauge boson, $B_{\mu}'$ is the $U(1)^{\prime}$ gauge boson and
$\chi$ is the kinetic mixing angle.
When the Higgs takes on a vev it potentially induces further mixing between these states.  The mass squared matrix is
\begin{equation}
{\mathcal M}_{Z}^{2}=
\left( \begin{array}{cc}
m_{Z_1}^2& \Delta_Z^2\\
\Delta_Z^2 &  m_{Z_2}^2\\
\end{array} \right)
\end{equation}
with
\begin{eqnarray}
m_{Z_1}^2&=&\frac{1}{4}\,g_Z^2\,v^2\\
m_{Z_2}^2&=&g'^2\,v^2(Q^{\prime\,\,\mathrm{mixed}}_{H_u}\,\cos^2{\beta}+
Q^{\prime\,\,\mathrm{mixed}}_{H_d}\,\sin^2{\beta})+(m_{Z^{\prime}}^{\mathrm{other}})^2\\
\Delta_Z^2&=&\frac{1}{2}\,g'\,g_Z\,v^2(Q^{\prime\,\,\mathrm{mixed}}_{H_u}\,\cos^
2{\beta}-Q^{\prime\,\,\mathrm{mixed}}_{H_d}\,\sin^2{\beta}).
\end{eqnarray}
Here, $g_Z^2=g_Y^2+g^2$, $\tan{\beta}=v_u/v_d$, $v^2=v_u^2+v_d^2$ and
$m_{Z^{\prime}}^{\mathrm{other}}$ parametrizes the contributions to the $Z^{\prime}$ mass from (exotic) SM singlets taking on vevs.  This results in the following mass eigenstates and $Z^0$-$Z^{\prime}$ mixing
angle:
\begin{eqnarray}
m^2_{Z^0,Z^{\prime}}&=&\frac{1}{2}\left(m_{Z_1}^2+m_{Z_2}^2\mp\sqrt{(m_{Z_1}^2-m_{Z_2}^2)^2+4\Delta_Z^4} \, \right)\\
   \tan{(2\theta_{Z^0\,Z^{\prime}})}&=&-2 \Delta_Z^2/(m_{Z_2}^2-m_{Z_1}^2).
\end{eqnarray}
The limit on $Z^0-Z^{\prime}$ mixing is $\theta_{Z^0 Z^{\prime}} <  \textrm{few} \times 10^{-3}$
\cite{Zprime_bound}.  While the precise value depends on model building details, it is not unreasonable to to take 
$\chi \sim 10^{-2}$. With this value and our choice of charges and parameters (Tables \ref{tab:ZprimeCH} and \ref{tab:bench1}), 
$\theta_{Z^0Z^{\prime}}=1.2\times10^{-3}$ ($\theta_{Z^0 Z^{\prime}}=1.9\times10^{-4}$) for a 1 TeV (2.5 TeV) $Z^{\prime}$
which satisfies this bound.  Note that we fix the mass of the $Z^{\prime}$ by hand (independent of $v_{s}$ and $g^{\prime}$), assuming
there are contributions from the additional physics contained within $m_{Z^{\prime}}^{\mathrm{other}}$.

\section{Collider signatures}\label{sec:Collider}
In this section, we consider the observability of the decay $Z^{\prime} \rightarrow$ Higgsinos.  The phenomenology of these decays will depend on the details of the supersymmetric particle spectrum.  We concentrate on the dramatic signal: $Z^{\prime} \rightarrow \tilde{\chi}_{i}^{0} \tilde{\chi}_{j}^{0} \rightarrow \tilde{\chi}_{1}^{0} \tilde{\chi}_{1}^{0} \ell^{+} \ell^{-} \ell^{+} \ell^{-}$, chosen for its particularly low SM and SUSY backgrounds.  Depending on the details of the superpartner spectrum, channels with hadronic activity might also be of use.  We will discuss two benchmark scenarios.  The first (section \ref{sec:onshellSlepton}) represents a particularly favorable case for the observation of $Z^{\prime} \rightarrow$ Higgsinos.  The second (section \ref{sec:onshellZ}) has a more generic spectrum,  but the desired decays of the $Z^{\prime}$ will be more challenging to observe.\footnote{There is also the possibility of studying gauge--mediated scenarios, where all SUSY events have some distinguishing feature: long-lived charged next-to-lightest supersymmetric particles (NLSPs), or photons coming from the decays of the NLSPs. In these cases it is clear that the SM backgrounds to our searches will be vanishing, and searching for the decays we discuss here should be straightforward.}

\begin{figure}[h]
\vspace{-.3in}
\begin{center}
    \includegraphics[width=.8\textwidth]{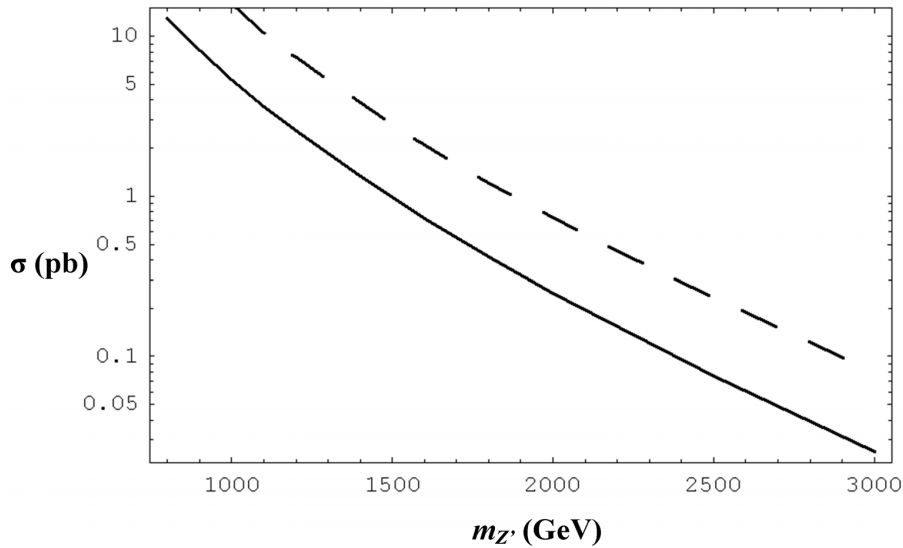}
\vspace{-.5in}
\end{center}
 \caption{Total cross section for $Z^{\prime}$ production at the LHC as a function of the $Z^{\prime}$ mass.  The solid line corresponds to a $Z^{\prime}$ with the benchmark charges of Table \ref{tab:ZprimeCH}.  For comparison, we have shown a dashed line for a sequential $Z^{\prime}$, a boson with charges and coupling identical to that of the SM $Z^0$.}
 \label{fig:ZPcross-section}

\end{figure}

In Fig \ref{fig:ZPcross-section} we have plotted the total $Z^{\prime}$ production cross section for the LHC for two different models using Pythia 6.4 \cite{Sjostrand:2006za} with the CTEQ 5L parton distribution functions\cite{Lai:1999wy}.  The solid line indicates the cross section used in this study (i.e. using the charges in Table \ref{tab:ZprimeCH}) with $g^{\prime}=0.6$.  The dashed line, shown for comparison purposes, is the production cross section for a sequential $Z^{\prime}$ with charges identical to those of the SM $Z^0$.  As expected, the cross section is a steeply falling function of the $Z^{\prime}$ mass.  This gives a rough indication the impact of $m_{Z^{\prime}}$ on the visibility of our measurement.
The rapid drop in cross section is somewhat mitigated by the presence of more $\displaystyle{\not}E_T$ and a harder lepton spectrum, since these effects increase the visibility of the events.  Unless otherwise noted, we set $m_{Z^{\prime}}=1$ TeV for the remainder of the paper.

\subsection{On-shell slepton}\label{sec:onshellSlepton}

We now examine a set of weak scale SUSY parameters that satisfies the following conditions: $\,\,m_{\tilde{\chi}_1^0} < m_{\tilde{\ell}}<
m_{\tilde{\chi}_{2,3}^0}\mathrm{,}\,\,m_{\tilde{\chi}_i^0}< m_{\tilde{q}}$ and $m_{\tilde{\chi}_{2,3}^0}-m_{\tilde{\chi}_1^0} < m_{Z^0}$.  These conditions ensure
$Z^{\prime}$ decays to $\tilde{\chi}_{i}^0+\tilde{\chi}_{j}^0$
will yield many $4\,\ell+\displaystyle{\not}E_T$ events via on-shell slepton decays for the neutralinos.  We postpone discussion of the case when the ``spoiler mode'' of the $\tilde{\chi}_{2}^{0} \rightarrow \tilde{\chi}_1^{0}\,Z^{0}$ is present until the next section.  Following these requirements we chose the parameters in Table \ref{tab:bench1}.  We take $M_2 =2 M_{1}$, motivated by unification of gaugino masses.

\begin{table}[h]
\begin{center}
\begin{tabular}{|l|l|}
\hline
$m_{Z^{\prime}} =1\,\mathrm{TeV}$ & $M_{1}=150 \,\mathrm{GeV}$ \\
$\Gamma_{Z^{\prime}}=15\,\mathrm{GeV}$ & $M_{2}=300\,\mathrm{GeV}(=2\times M_{1})$\\
$g' =0.6$ & $\mu=200 \,\mathrm{GeV}$ \\
$m_{\tilde{\ell}} =160\,\mathrm{GeV}$ & $\tan{\beta}=5$\\
$m_{\tilde{q}}  > 1000\,\mathrm{GeV}$ &  $m_{\mathrm{exotics}} > m_{Z^{\prime}}/2$\\
\hline\end{tabular}
\end{center}
\caption{The weak scale parameters relevant for the on-shell slepton study.  We take the $Z^{\prime}$ charges to be as in Table \ref{tab:ZprimeCH}
}
\label{tab:bench1}
\end{table}

We also choose parameters so that the neutralino mixing matrix is block diagonal -- the four lightest states are MSSM-like while the heavier two are a mixture of the singlino and $Z^{\prime}$-ino (see \cite{HaberChoi} for details about the neutralino phenomenology when this approximation does not hold).   The dominant contribution to the four lepton signal comes comes from the production and decay of $\tilde{\chi}_{2}^0$.  It is an almost equal admixture of $\tilde{B^0},\,\,\tilde{W^3},\,\,\tilde{H_d^0},\,\,\tilde{H_u^0}$.  This composition leads a BR($Z^{\prime} \rightarrow \tilde{\chi}_{2}^{0} \tilde{\chi}_{2}^{0}) = 0.9\% $ (via the Higgsino content) and BR($\tilde{\chi}_{2}^{0} \rightarrow \tilde{\chi}_1^0$ + leptons)$\approx 65\%$ (via its bino and wino content).  The four lepton signal is suppressed for $\tilde{\chi}_{3}^0$ and $\tilde{\chi}_{4}^0$ due to the small bino and wino content of the former and the small Higgsino content of the latter. The neutralino masses are given in Table \ref{tab:mchiOnshellSlepton}.
\begin{table}[h]
\begin{center}
\begin{tabular}{|c|c|c|c|c|c|c|}
\hline
$i$ & 1 & 2 & 3 & 4 & 5 & 6\\
\hline
$m_{\tilde{\chi}_i^0}$ (GeV) & 126 & 192 & 206 & 338 & $>m_{Z^{\prime}}/2$ & $>m_{Z^{\prime}}/2$ \\
\hline
\end{tabular}
\end{center}
\caption{Neutralino masses for on-shell slepton study.  Parameters are given in Table \ref{tab:bench1}.
}
\label{tab:mchiOnshellSlepton}
\end{table}

Following the study done in \cite{CMSNote}, the dominant SM backgrounds to $4 \ell+\displaystyle{\not}E_T$ are due to $t\,\overline{t}$, $Z^0\,b\,\overline{b}$ and $Z^0\, Z^0$ production.  A jet veto effectively eliminates the two colored modes, justifying our choice to concentrate on the background of $Z^0\, Z^0$ production.

Similarly, after a jet veto the most relevant MSSM background is direct neutralino production via an off-shell $Z^0$ where the neutralinos then decay leptonically.  We will refer to this type of production as continuum production.  Other contributions, coming from cascade decays of squarks and gluinos, are sub-dominant after the jet veto.  The precise contribution depends on the details of the squark and gluino spectrum.

There can also be contributions to the continuum background from t-channel squark exchange.  Depending on the squark mass, this can actually increase or decrease the neutralino production via interference.  We neglect this diagram for our study, taking the limit where the squarks are heavy.  There is also a potential contribution to neutralino pair production background via an s-channel heavy Higgs ($A^0$ and $H^{0}$).  For the present discussion, we make the conservative assumption that mixing between the MSSM Higgs bosons and any new scalars from the singlet sector are small.  After cuts, $m_{A^0} \approx 800$ GeV gave the largest cross section.  This mass balances falling production cross section against an increasing likelihood to pass the relevant cuts.   Even at this mass, however, the contribution to the background was still sub-dominant to direct neutralino production through a $Z^{0*}$, contributing only about 30$\%$ of the continuum background.  Bearing in mind the possibility of additional (small) contributions to the MSSM background, in what follows we focus on $Z^{0*}$ mediated production-- the one contribution that must be there -- and in any case is usually dominant.

Before cuts Pythia gives a continuum cross section
\begin{equation}
\sum_{\ell_1,\ell_2,i,j}\sigma(p\,p\rightarrow\tilde{\chi}_i^0\,\tilde{\chi}_j^0)\,\mathrm{BR}(\tilde{\chi}_i^0\rightarrow \ell_1^+\,\ell_1^-\,\displaystyle{\not}E_T)\,\mathrm{BR}(\tilde{\chi}_j^0\rightarrow \ell_2^+\,\ell_2^-\,\displaystyle{\not}E_T)\approx 8.8\,\,\mathrm{fb}.
\end{equation}
This result should be contrasted with the resonant production.  Pythia is capable of producing on-shell $Z^{\prime}$'s but does not
decay them to MSSM particles.  So, we used Pythia to calculate
$\sigma(p\,p\rightarrow Z^{\prime})$, and used the appropriate BRs to calculate
\begin{equation}
\sum_{l_1,l_2}\sigma(p\,p\rightarrow Z^{\prime})\times\mathrm{BR}(Z^{\prime}\rightarrow \ell_1^+\,\ell_1^-\,\ell_2^+\,
\ell_2^-\,\displaystyle{\not}E_T) \approx
\left( \begin{array}{cccc}
 0 & 0 & 0  & 0 \\
 0 & 20 & 0.73 & 4.0\\
 0 & 0 & 3.3 & 2.6\times10^{-5}\\
 0 & 0 & 0 & 0.18 \\
 \end{array} \right)  \rm fb.
\label{eq:onshellslxsec}
\end{equation}
Here the rows ($i$) and columns ($j$) refer to contributions from $Z^{\prime}\rightarrow\tilde{\chi}_i^0\,\,\tilde{\chi}_j^0$.  We then modeled the $Z^{\prime}$ resonance by producing neutralino pairs with $m_{Z^{\prime}}-\Gamma_{Z^{\prime}} < \sqrt{s} < m_{Z^{\prime}}+\Gamma_{Z^{\prime}}$ and then scaled the cross section according to Eqn. (\ref{eq:onshellslxsec}).

\begin{figure}[h]
\vspace{-.3in}
\begin{center}
    \includegraphics[angle=270,width=.8\textwidth]{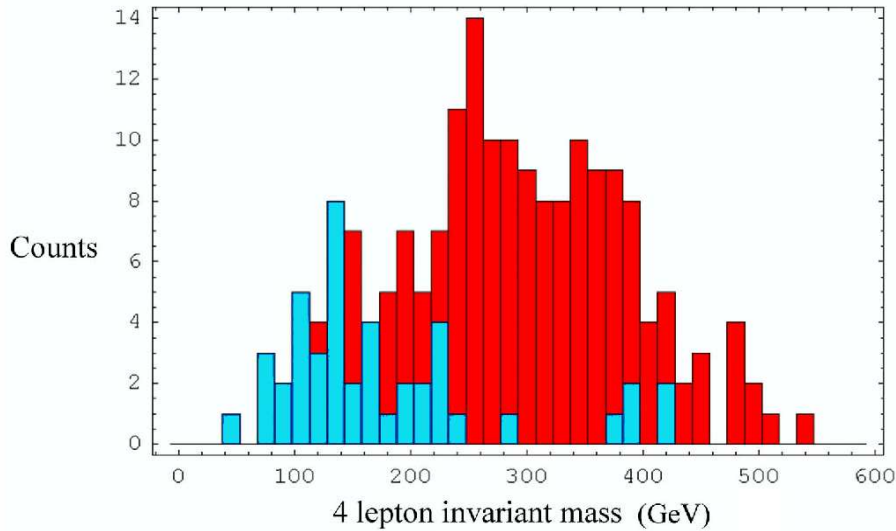}
\end{center}
\vspace{-.5in}
 \caption{A histogram of the invariant mass of the four leptons in the events selected.  All cuts except the invariant mass cut have been applied.  Displayed are the continuum background (light blue) and the signal coming from the $Z^{\prime}$ decay (dark red).  The data shown corresponds to 30 fb$^{-1}$ for a 1 TeV $Z^{\prime}$ with charges as in Table \ref{tab:ZprimeCH}.}
 \label{fig:InvMass}
\end{figure}

 These properly scaled results were then piped through the PGS simulation\cite{PGS4} to account for simple detector effects.  Events where two sets of opposite sign, same flavor (OSSF) leptons were detected were selected.  No event was allowed to have a jet with $p_T>30$ GeV.  Finally events were required to have $\displaystyle{\not}E_T>50$ GeV and the invariant mass of the four leptons greater then 300 GeV.  After the jet veto, the $\displaystyle{\not}E_T$ cut effectively eliminated the remaining SM background. The invariant mass cut greatly reduced the remaining MSSM background, see Fig.~\ref{fig:InvMass}. The $\displaystyle{\not}E_T$ and invariant mass cuts were chosen to maximize $\mathrm{signal}/\sqrt{\mathrm{background}}$.

The post-cut results are shown in Table \ref{tab:resultonshellSlepton}.  With 30 fb$^{-1}$ of integrated luminosity and a 1 TeV $Z^{\prime}$ we can expect to see $\approx 90$ events with only a handful of background events.  Even for 10 fb$^{-1}$ we should be able to claim a discovery.

To see how this signal depends on the mass of the $Z^{\prime}$, we repeated the above process for $m_{Z^{\prime}}=2.5$ TeV.  
We (now more optimistically) assumed that the exotics were still too heavy to contribute.  The only cut that changed was the invariant mass cut, now taken at 500 GeV.  This took advantage of the harder spectrum for the signal leptons coming from the heavier resonance. After cuts a signal cross section of  $\approx 0.08$ fb remained with a vanishing background.  Hence we only expect $\approx 2$ events with 30 fb$^{-1}$.  Since the background is even smaller for this case (due to the larger invariant mass cut) we would be able to see a signal with 100 fb$^{-1}$.  However, we expect the jet veto will be less effective once the LHC begins running at higher luminosity.

\begin{table}[h]
\begin{center}
\begin{tabular}{|c|c|c|c|}
\hline
       $m_{Z^{\prime}}=1$ TeV     (all entries in fb)         & SM (diboson)   & continuum & $Z^{\prime}$\\
\hline
$\sigma\times\epsilon$                    & $23.2 \pm 0.1$ & $4.2\pm0.1$       & $14.4\pm0.5$ \\
jet veto $(p_T \ngtr 30$ GeV)             & $18.5 \pm 0.1$ & $3.0\pm0.1$      & $8.8\pm0.4$\\
$\displaystyle{\not}E_T > 50$ GeV         & $0.025 \pm 0.004$ & $1.50\pm 0.04$ & $6.1\pm0.3$\\
Invariant Mass of 4 $\ell > 300$  GeV     & $0.004 \pm 0.002$ & $0.13\pm0.01$ & $3.0\pm0.2$ \\
\hline
\end{tabular}
\end{center}
\caption{Results for on-shell slepton study.  SM refers to $Z^0\,\,Z^0$ production, continuum refers to $Z^{0*}\rightarrow\tilde{\chi}_i^0\,\,\tilde{\chi}_j^0$.  $\epsilon$ is the detector efficiency times the BR for 4 $\ell+\displaystyle{\not}E_T$ for each process.  The errors shown are statistical in nature, due to a limited number of simulation events.}
\label{tab:resultonshellSlepton}
\end{table}

\subsection{On-shell $Z^0$}\label{sec:onshellZ}
The above study required a fortuitous mass spectrum.  What would happen if the spectrum were not as favorable?  If the splitting between the $\chi_2^{0}$ and $\chi_{1}^{0}$ is sufficiently large, then the neutralinos dominantly decay via an on-shell $Z^0$.  The small branching fraction of $Z^0\rightarrow \ell^+\,\ell^-$ causes the BR($\tilde{\chi}_i^0\rightarrow\ell^+\ell^-\tilde{\chi}_1^0$) to be greatly reduced when compared with the on-shell slepton study.  The parameters for the on-shell $Z^0$ study are given in Table \ref{tab:bench2}.  
\begin{table}[h]
\begin{center}
\begin{tabular}{|l|l|}
\hline
                   $m_{Z^{\prime}}=1\,\mathrm{TeV}$          & $ M_{1}=150\,\mathrm{GeV}$\\
                   $\Gamma_{Z^{\prime}}=13\,\mathrm{GeV}$   & $M_{2}=300\,\mathrm{GeV}\,\,(=2\times M_{1})$ \\
                    $ g'=0.6$                       & $ \mu=300\,\mathrm{GeV}$ \\
                   $m_{\tilde{l}}=500\,\mathrm{GeV}$  & $\tan{\beta}=5$\\
                  $ m_{\tilde{q}}>1000\,\mathrm{GeV}$ & $ m_{\mathrm{exotics}}>m_{Z^{\prime}}/2$\\
\hline
\end{tabular}
\end{center}
\caption{Weak scale parameters for the on-shell $Z^0$ study.   Again, we have assumed that all exotics are sufficiently heavy that decays from the $Z^{\prime}$ are inaccessible. The widths are calculated under this assumption using the charges of Table \ref{tab:ZprimeCH}.}
\label{tab:bench2}
\end{table}

Again we chose the  neutralino matrix to be block diagonal so we will ignore the singlino and $Z^{\prime}$-ino contributions.  The biggest contribution to the four lepton signal comes from $\tilde{\chi}_{2}^0$ and $\tilde{\chi}_{3}^0$. 
For $\tilde{\chi}_{2,3}^0$ the only kinematically allowed decays are  to $Z^0\,\tilde{\chi}_{1}^0$ so their BR to leptons is approximately equal to BR($Z^0\rightarrow\ell^+\,\ell^-$).  The fact that they have sizable Higgsino content guarantees that they will be produced in abundance.  The $\tilde{\chi}_{4}^0$ decays almost always to $\tilde{\chi}^+_1  W^- \rightarrow W^+ W^-  \tilde{\chi}^0_1$.  The neutralino masses for the on-shell $Z^0$ study are given in Table \ref{tab:mchionshellZ}.

\begin{table}[h]
\begin{center}
\begin{tabular}{|c|c|c|c|c|c|c|}
\hline
 $i$ & 1 & 2 & 3 & 4 & 5 & 6\\
\hline
 $m_{\chi^0_i}$ (GeV) & 142 &242 & 305 & 371 & $>m_{Z^{\prime}}/2$ & $>m_{Z^{\prime}}/2$\\
\hline
\end{tabular}
\end{center}
\caption{Neutralino masses for the on-shell $Z^0$ study.  Parameters are given in Table \ref{tab:bench2}.
}
\label{tab:mchionshellZ}
\end{table}

The analysis proceeds as in the previous section.  We select events with 2 sets of OSSF leptons.   After the jet veto (again rejecting events with jet $p_T>30$ GeV) the dominant SM background is $Z^0\,Z^0$ production. For the continuum (again neglecting the squark and heavy Higgs contributions) we used Pythia to calculate (before cuts)
\begin{equation}
\sum_{\ell_1,\ell_2,i,j}\sigma(p\,p\rightarrow\tilde{\chi}_i^0\,\tilde{\chi}_j^0)\,\mathrm{BR}(\tilde{\chi}_i^0\rightarrow \ell_1^+\,\ell_1^-\,\displaystyle{\not}E_T)\,\mathrm{BR}(\tilde{\chi}_j^0\rightarrow \ell_2^+\,\ell_2^-\,\displaystyle{\not}E_T)\approx 0.16\,\,\mathrm{fb}.
\end{equation}

The same combination of Pythia and analytics discussed in section \ref{sec:onshellSlepton} gives
\begin{equation}
\sum_{\ell_1,\ell_2}\sigma(p\,p\rightarrow Z^{\prime})\times\mathrm{BR}(Z^{\prime}\rightarrow \ell_1^+\,\ell_1^-\,\ell_2^+\,
\ell_2^-\,\displaystyle{\not}E_T) \approx
\left( \begin{array}{cccc}
 0 & 0 & 0  & 0 \\
 0 & 0.19 & 6.9\times10^{-2} & 0.19\\
 0 & 0 & 0.61 & 9.9\times10^{-3}\\
 0 & 0 & 0 & 4.3\times10^{-2} \\
 \end{array} \right)  \rm fb.
\label{eq:ZPxsection}
\end{equation}
Here the rows ($i$) and columns ($j$) refer to the contribution from $Z^{\prime} \rightarrow\tilde{\chi}_i^0\,\,\tilde{\chi}_j^0$.

Table \ref{tab:OnShellZResults} shows that after cuts the resultant signal cross section is more then an order of magnitude larger then the backgrounds.  However, its small size presents a challenge since there will be only $\approx 7$ events for 30 fb$^{-1}$ of integrated luminosity.  Although the backgrounds will produce at most 1 event for this amount of data, one might worry that this low number of events would not be enough to claim discovery.  At higher luminosities,when pile-up can be significant, careful studies will need to be done to test the efficacy of the jet veto.

The final state $Z^{\prime} \rightarrow 2 \ell + \displaystyle{\not}E_T$ also holds some promise in this case.  The signal is less distinctive; hence SM backgrounds become an issue.  However, the higher branching ratio for $Z \rightarrow \nu \bar{\nu}$ allows for an increased rate.  Our studies indicate that it is not as useful as the four lepton final state for the benchmark considered here, but it could be considered as a complementary analysis, depending on the supersymmetric spectrum that nature chooses.

\begin{table}[h]
\begin{center}
\begin{tabular}{|c|c|c|c|}
\hline
$m_{Z^{\prime}}=1$ TeV (all entries in fb) & SM (diboson) & continuum & $Z^{\prime}$\\
\hline
$\sigma\times\epsilon$                               & $23.2 \pm 0.1$     & $0.089\pm0.0012$  & $0.64\pm 0.02$ \\
jet veto $(p_T \ngtr 30$ GeV)                       & $18.5 \pm 0.1$   & $0.060\pm0.001$   & $0.38\pm 0.01$\\
$\displaystyle{\not}E_T > 40$                       & $0.041 \pm 0.005$ & $0.055\pm0.001$  & $0.33\pm 0.01$\\
Invariant Mass of $4 \ell >$ 300 GeV               & $0.005 \pm 0.002$ & $0.020\pm0.001$ & $0.24\pm 0.01$ \\
\hline
\end{tabular}
\end{center}
\caption{Results for the on-shell $Z^0$ study (see Table \ref{tab:bench2} for explicit parameters).  SM refers to $Z^0\,\,Z^0$ production, while continuum refers to $Z^{0*}\rightarrow\tilde{\chi}_i^0\,\,\tilde{\chi}_j^0$.  The final column gives $Z^{\prime}\rightarrow\tilde{\chi}_i^0\,\,\tilde{\chi}_j^0$. $\epsilon$ is the detector efficiency times the BR for 4 $l+\displaystyle{\not}E_T$ for each process.  The errors shown are statistical in nature, due to a limited number of simulation events.}
\label{tab:OnShellZResults}
\end{table}

\section{Are we really forbidding the $\mu$-term?}\label{sec:Implications}
Once the $Z^{\prime}\rightarrow\tilde{\chi}_i^0\,\tilde{\chi}_j^0$ signal has been observed, one might suspect that the new gauge symmetry is responsible for forbidding a bare $\mu$-term.  How can we solidify this conclusion?  After all, having  the Higgsinos charged under the $U(1)^{\prime}$ is only a necessary condition to forbid $\mu$.  It is not sufficient; it is possible to have $Q_{H_{u}} = - Q_{H_{d}} \neq 0$.  In the following we refer to $Q'_{H_d}=-Q'_{H_u}$ as $\mu$-allowed and $Q'_{H_d}\ne -Q'_{H_u}$ as $\mu$-forbidden.  

One approach to test whether the $\mu$-term is forbidden is independent of the neutralino spectrum.   Since the MSSM the superpotential must contain $Q\,U\,H_u$ and $Q\,D\,H_d$,  the $U(1)^{\prime}$ charges must satisfy
\begin{eqnarray}
Q'_Q+Q'_U+Q'_{H_u}&=&0\\
Q'_Q+Q'_D+Q'_{H_d}&=&0
\label{eq:YukawaConditions}
\end{eqnarray}
So for $\mu$-allowed:
\begin{equation}
\frac{Q'_U}{Q'_Q}+\frac{Q'_D}{Q'_Q}=-2.
\label{eq:chargeCondition}
\end{equation}
Violations of this equality would be an indication that the $U(1)^{\prime}$ was forbidding the $\mu$-term.  Previous studies (\cite{delAguila:1993ym},\cite{ZPcouplings}) considered $Z^{\prime}$ observables such as the forward-backward asymmetry and detailed rapidity distributions in $Z^{\prime} \rightarrow \ell^{+} \ell^{-}$ final states, along with various rare decays.  In \cite{delAguila:1993ym} it was determined that for a 1 TeV $Z^{\prime}$ and 100 fb$^{-1}$, $(Q'_U/Q'_Q)^2$ and $(Q'_D/Q'_Q)^2$ could be determined within about $20\%$ for the former and a range of errors from $7\%$ to more then $100\%$ for the later, depending on the choice of model.  These measurements are only sensitive to the squares of the charges, and hence not their sign.  This leads to an ambiguity in testing Eq. (\ref{eq:chargeCondition}).  For brevity, we square this equation while leaving the sign undetermined as follows:
\begin{equation}
\left(\frac{Q'_U}{Q'_Q}\right)^2+\left(\frac{Q'_D}{Q'_Q}\right)^2\pm2 \left|\frac{Q'_U}{Q'_Q}\right|\left|\frac{Q'_D}{Q'_Q}\right|-4=0.
\label{eq:chargeConditionSquared}
\end{equation}
Now we apply Eqn. (\ref{eq:chargeConditionSquared}) to the results of \cite{delAguila:1993ym}.  Since we are unable to measure the relative signs we must try both.  If either choice results in the condition being satisfied, we are left with an indeterminate result.  However, if Eqn. (\ref{eq:chargeConditionSquared}) cannot be satisfied we can be certain that the $\mu$-term is forbidden.

To get a feel for how well this technique works we examine several $U(1)^{\prime}$ models.  We consider the four models studied in  \cite{delAguila:1993ym}, along with the model considered in the previous sections, which will denote as $N$.  By extrapolation from the charges of the other models, we make a rough error estimate of $20\%$ for $(Q'_U/Q'_Q)^2$ and $30\%$ for $(Q'_D/Q'_Q)^2$ for the $N$ model.  For the other models, we take the error estimates directly from \cite{delAguila:1993ym}.  A naive combination of errors leads to the determinations of Table \ref{table:chargeDetermination}.  The ``+" and ``-" columns indicate the result of the left hand side of Eqn. (\ref{eq:chargeConditionSquared}) based on the choice of sign.  Deviations from 0 indicate that the $\mu$-term is forbidden.  From Table \ref{table:chargeDetermination}, one can see that in 3 of the 5 cases these observables are not enough to probe the status of the $\mu$-term.  Even for the $N$ and $\eta$ models there is little more than 2$\sigma$ confidence that the $\mu$-term is forbidden.  Obviously we need further observables to resolve these ambiguities.

\begin{table}[h]
\begin{center}
\begin{tabular}{|c|c|c|c|c|c|}
\hline
 Model & $+$ & $-$ & uncertainty& theoretical & experimental\\
\hline
         $N$ &   5     &    -3 & 1.4       & forbidden      &  forbidden          \\
         $\chi$  & 12 & 0 & 0.9           & allowed &       indeterminate\\
         $\psi$   &  0  & -4 & 0.7         & forbidden    &   indeterminate\\
         $\eta$    & -1.8   & -3.8 & 0.7 & forbidden & forbidden\\
         LR      & 196   & 0 & 22  &  allowed &  indeterminate\\
\hline
\end{tabular}
\end{center}
\caption{Results for applying  Eqn. (\ref{eq:chargeConditionSquared}) to different $Z^{\prime}$ models.  ``+" corresponds to taking $\mathrm{sgn}(Q'_U/Q'_Q)=\mathrm{sgn}(Q'_U/Q'_Q)$ and ``-" corresponds to taking $\mathrm{sgn}(Q'_U/Q'_Q)\ne\mathrm{sgn}(Q'_U/Q'_Q)$.  The uncertainty corresponds to how well we can determine these ratios at the LHC with 100 fb$^{-1}$ for a 1 TeV $Z^{\prime}$.  The theoretical column refers to the status of the $\mu$-term for the specific charges of each model.  The experimental column refers to the determination we can make (at 2$\sigma$) using the data of \cite{delAguila:1993ym}.  A non-zero result for both ``+" and ``-" is equivalent to $\mu$-forbidden.}
\label{table:chargeDetermination}
\end{table}

One such complementary analysis would be to apply the Dalitz plot-like wedge-box technique of \cite{Wedgebox} to the four lepton events studied in the previous sections.  The idea is to pair leptons from the same parent neutralino (perhaps by only using events with a pair of opposite sign electrons and opposite sign muons).  One then plots the invariant mass of the first pair against the invariant mass of the second pair.  Assuming the neutralino mass splittings are less than $m_{{Z}^{0}}$, kinematic endpoints will lead to a ``box" shape for the case where the parent particles have the same mass and a ``wedge" shape when their masses are different.  In a box plot the majority of events lie within a square; a wedge shape occurs when the events lie within two perpendicular rectangles.  With enough statistics, this technique can tell us if the dominant neutralino production is mostly due to $Z^{\prime}\rightarrow \tilde{\chi}_i^0\,\tilde{\chi}_i^0$ events (diagonal production) or $Z^{\prime}\rightarrow \tilde{\chi}_i^0\,\tilde{\chi}_j^0,\,\,i\ne j$ events (off-diagonal production).

A complication occurs when both neutralinos can decay to an on-shell $Z^0$.  In this case, the identity of the parent neutralino is no longer encoded in the invariant mass of the sleptons --- they simply reconstruct a $Z^0$.  One can instead examine the $p_{T}$ spectrum of the reconstructed $Z^0$.  Those with larger $p_{T}$ come from the heavier neutralinos.  Then, in principle, one could form a wedge-box plot of the two $Z^0$ boson $p_{T}$'s.  In practice, however, the event sample of four lepton events is probably too small, at least with 30 fb$^{-1}$.

To understand in detail why this technique is useful for determining the status of the $\mu$-term, recall that in the absence of $Z^0-Z^{\prime}$ mixing the $Z^{\prime}\rightarrow\tilde{\chi}_i^0\,\tilde{\chi}_j^0$ vertex is proportional to $(Q'_{H_d}\,N_{i,H_d}\,N_{j,H_d}+Q'_{H_u}\,N_{i,H_u}\,N_{j,H_u})$.  $N$ is the neutralino mixing matrix which we take to be real.  In the $\mu$-allowed case this reduces to $Q'_{H_d}(N_{i,H_d}\,N_{j,H_d}-N_{i,H_u}\,N_{j,H_u})$.  This has the same form as the neutralino-$Z^0$ coupling in the MSSM.   Off-diagonal production dominates in this case (see for example \cite{Haber:1984rc}) since $i=j$ vertices will always suffer some degree of cancellation.  To understand this effect take the limit of pure Higgsinos: $\mu> M_{2} > M_{1}\gg m_{Z^0}$ and neglect the additional singlino and $Z^{\prime}$-ino states.  Then the neutralino mixing matrix is given by the approximately block diagonal form:
\begin{equation}
N =\left( \begin{array}{cccc}
1 & 0 & 0 & 0 \\
0 & 1 & 0 & 0 \\
0 & 0& \sqrt{2}/2 & -\sqrt{2}/2\\
0 & 0& \sqrt{2}/2 & \sqrt{2}/2  \\
 \end{array} \right)
\end{equation}
Noting that the pure Higgsino states are $\tilde{\chi}_3^0$ and $\tilde{\chi}_4^0$, $Z^{\prime}\nrightarrow\tilde{\chi}_i^0\,\tilde{\chi}_i^0$ while the $Z^{\prime}\rightarrow\tilde{\chi}_3^0\,\tilde{\chi}_4^0$ vertex survives as $\mathrm{sgn}(N_{3,H_d}\times N_{4,H_d})\ne \mathrm{sgn}(N_{3,H_u}\times N_{4,H_u})$. The dominance of off-diagonal production manifests as a wedge when one performs a wedge-box analysis (see Fig \ref{fig:wedge-allowed}).

\begin{figure}[h]
  \begin{center}
  \includegraphics[width=.8\textwidth]{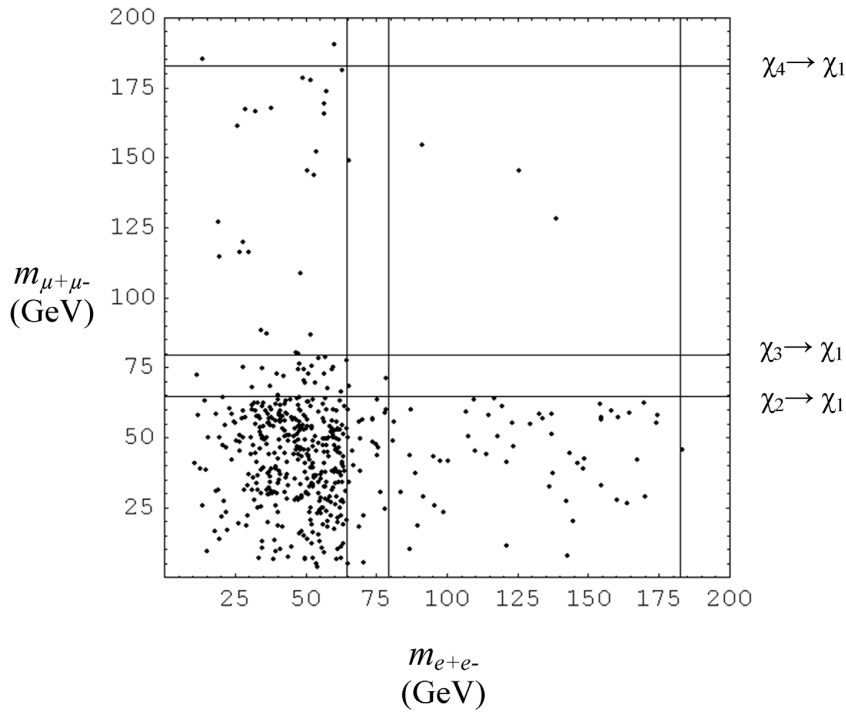}
  \end{center}
  \caption{Wedge-box plot for the case where the $\mu$-term is forbidden by the $U(1)^{\prime}$ symmetry.  The SUSY parameter choices are those taken in Sec. \ref{sec:onshellSlepton}. We have plotted 4 lepton events with a pair of OS electrons and OS muons from $Z^{\prime}$ resonance neutralino decays.  The x-axis and y-axis correspond to the invariant mass of the electron pair and the muon pair respectively.  The charges can be found in Table \ref{tab:ZprimeCH}.  We have plotted 500 points for illustration.  The solid lines (labeled $\chi_i\rightarrow \chi_1$) correspond to the expected kinematic edges for the masses given in Table \ref{tab:mchiOnshellSlepton}.  Note the density of points in the lower left corner corresponding to large $Z^{\prime}\rightarrow\tilde{\chi}_i^0\,\,\tilde{\chi}_j^0$ production.  This is an example of a ``box."}
  \label{fig:wedge-forbidden}
\end{figure}

\begin{figure}[h]
  \begin{center}
  \includegraphics[width=.8\textwidth]{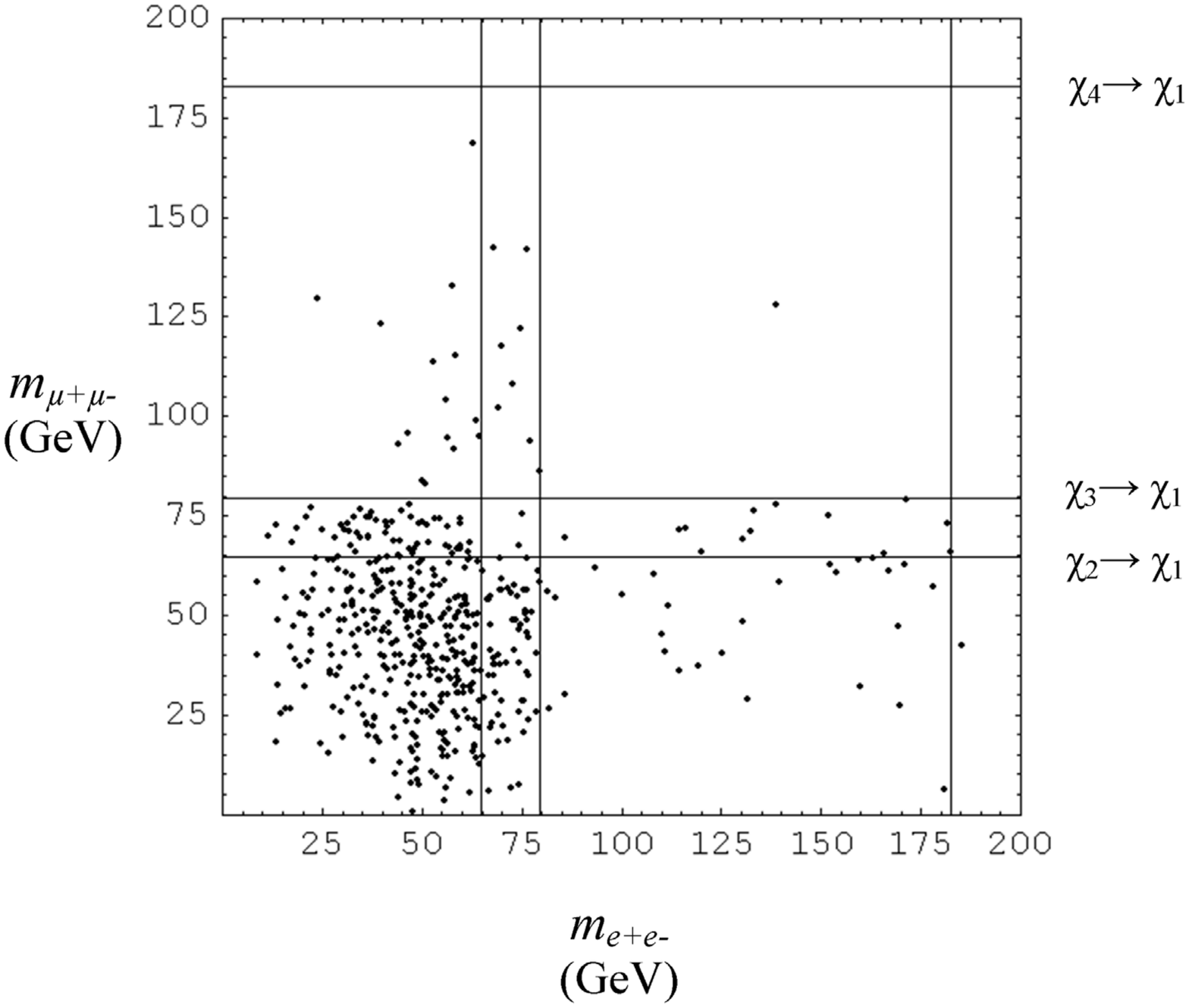}
  \end{center}
  \caption{Wedge-box plot is for the case where the $\mu$-term is allowed by the $U(1)^{\prime}$ symmetry,  i.e. $Q'_{H_u}=-Q'_{H_d}$.  See Figure \ref{fig:wedge-forbidden} for a detailed explanation.  The density of the points is relatively uniform out to the $\chi_{3} \rightarrow \chi_1$ line, excluding the region where both invariant masses are beyond the $\chi_{2} \rightarrow \chi_{1}$ line. This indicates sizable off-diagonal ($\tilde{\chi}_{3}^{0} \tilde{\chi}_{2}^{0}$) production.  This is what makes a wedge a wedge. }
  \label{fig:wedge-allowed}
\end{figure}

As an example of the wedge vs. box effect we looked the fraction of $4\,\ell+\displaystyle{\not}E_T$ events due to $Z^{\prime}\rightarrow\tilde{\chi}_i^0\,\,\tilde{\chi}_j^0$ for different values of $i$ and $j$ where the SUSY parameters are those of Sec. \ref{sec:onshellSlepton}.  The results are displayed in Table \ref{tab:diagVSoffdiag} where we can see that for $\mu$-allowed, diagonal production accounts for 0.2$\%$ of total events as opposed to $\mu$-forbidden where diagonal production makes up about 83$\%$.  These differences in the cross sections for different production channels would show up in the wedgebox plots as shown in Figures \ref{fig:wedge-forbidden} and \ref{fig:wedge-allowed}.  The wedge-box technique could therefore provide a powerful window into the fundamental nature of the $\mu$-term.

\begin{table}[h]
\begin{center}
\begin{tabular}{|c|c|c|c|c|c|c|}
\hline
$\mu$ & $(2,2)$ & $(3,3)$ & $(4,4)$ & $(2,3)$ & $(2,4)$ & $(3,4)$\\
\hline
allowed  & $7.0\times 10^{-4}$ & $2.7\times 10^{-4}$ & $9.7\times 10^{-4}$ &  $8.8\times 10^{-1}$ &  $4.4\times 10^{-3}$ &  $1.2\times 10^{-1}$\\
\hline
forbidden  & $7.1\times 10^{-1}$ & $1.2\times 10^{-1}$ & $6.5\times 10^{-3}$ &  $2.6\times 10^{-2}$ &  $1.4\times 10^{-1}$ &  $9.2\times 10^{-7}$\\
\hline
\end{tabular}
\end{center}
\caption{Fraction of $Z^{\prime}\rightarrow\tilde{\chi}_i^0\,\,\tilde{\chi}_j^0\rightarrow4\,\ell\,\displaystyle{\not}E_T$ events due to different neutralino production channels.  The SUSY parameters are given in Table \ref{tab:bench1}.  The ordered pair correspond to the neutralino pair $(i,j)$.  For $\mu$-forbidden we use the choice of charges stated in Table \ref{tab:ZprimeCH}.
}
\label{tab:diagVSoffdiag}
\end{table}

The applicability of this method depends on the details of the SUSY spectrum.  The most pressing issue is that the wedge-box plot is created from the daughter leptons, not the neutralinos themselves.  Therefore, there is a danger that a box or wedge shape could be a reflection of differences in the branching ratio to leptons, rather than the production cross section from the $Z^{\prime}$ decays.    Typically, however, the cancellation of the diagonal production is very effective in the case where the $\mu$-term is allowed.  Thus, the branching ratio to sleptons must be different by large factors to turn  wedge-like plot into a box.  For neutralino parameters similar to those considered in section \ref{sec:onshellSlepton} (but for $\mu$-allowed), the branching ratio of the relevant neutralinos to leptons would have to differ by roughly two orders of magnitude in order to become a box.  So, we can view the presence of a box as strong evidence that the $\mu$-term is forbidden in spite of this complication.  In addition, it is not unreasonable to expect that one could learn about neutralino branching ratios from other samples of events, e.g., cascade decays, and thereby illuminating this issue.

We now comment on the robustness of the conclusions that one can draw from these plots. It is relatively straightforward to get a wedge even if the $\mu$-term is forbidden.  As a trivial example, note $Q'_{H_d}=-(1+\epsilon)Q'_{H_u}$ with $\epsilon$ small forbids the $\mu$-term but the $Z^{\prime}$ dominantly decays to off-diagonal neutralino pairs. So, while the observation of a wedge-like plot does not say anything definitive about the status of the $\mu$-term, a ``box''-like plot is a strong indicator that the $\mu$-term is forbidden the gauge symmetry.

We conclude this section with a brief discussion of two other observables that seem difficult to measure, particularly at the LHC, but potentially provide insight.  For example, one might attempt to measure decays of the type $Z^{\prime} \rightarrow H_{i}\,A^{0}$ or $Z^{\prime} \rightarrow h_{i}\,Z^{0}$. The relative branching ratios encode information about the $U(1)^{\prime}$ charges of the Higgs multiplets.  However, while the observation of  $Z^{\prime} \rightarrow h_{i}\,Z^{0}$ seems feasible, (at least for the lightest Higgs-- this was recently studied in the context of Little Higgs models \cite{LH, LH2}), the other channels seem more difficult.  Furthermore, the extraction of information also depends on how close the Higgs sector is to the decoupling limit. There is also the potential that mixing with the singlets of the $Z^{\prime}$ sector could complicate the phenomenology.   Another potential observable is the angular dependence of the charginos in $Z^{\prime}$ decays.  The $Z^{\prime}\rightarrow\tilde{\chi}_i^+\,\tilde{\chi}_j^-$ vertex is proportional to $\gamma^{\mu}(c_V-c_A\gamma_5)$ where $(c_V)_{i,j}=Q'_{H_u}V_{i,2}V_{j,2}-Q'_{H_d}U_{i,2}U_{j,2}$ and $(c_A)_{i,j}=Q'_{H_u}V_{i,2}V_{j,2}+Q'_{H_d}U_{i,2}U_{j,2}$.  An $A_{FB}$ measurement could be used to determine $c_V$ and $c_A$.  When coupled with information about the chargino mixing matrices, this would give the charges of $H_u$ and $H_d$ under the $U(1)^{\prime}$.  Of course this would require isolating a sample of chargino decays, determining which specific charginos were being observed and then doing detailed measurements of their angular distributions.  While this would be a difficult task, in principle this measurement could also tell us about the $\mu$-term, providing a consistency check with the Higgsino and/or Higgs measurements.

\section{Conclusions}

If a gauge symmetry is responsible for forbidding the $\mu$-term, it is possible that one might be able to observe decays of the type $Z^{\prime} \rightarrow$ Higgsinos, via leptonic decays of the neutralinos.  The ease with which this signal will be seen depends sensitively on the superparticle spectrum.  If the signal is observed, kinematic information in the decays might be sufficient to determine definitively whether the $U(1)^{\prime}$ forbids the $\mu$-term.  Other complementary approaches, involving measurement of the quark charges, or examining $Z^{\prime}$ decays to Higgs bosons and/or charginos might strengthen these conclusions.

While in this paper we were primarily concerned with probing the Higgsino charges under the new gauge symmetry, the $Z^{\prime}$ potentially has another use.  It presents a new source of Higgsinos at the LHC, beyond those available in direct production and cascade decays.  For example, in Table \ref{tab:resultonshellSlepton}, we can see that the production of the Higgsinos via the $Z^{\prime}$ can far exceed direct production.  (A similar point was made for sleptons in \cite{Baumgart}.) The $Z^{\prime}$ can help us study parts of the SUSY spectrum that might not otherwise be readily accessible.

\section*{Acknowledgments}
We thank Gordon Kane and James Wells for useful comments.  The work of AP is supported by the University of Michigan and the MCTP.  The work of TC is supported in part by the DOE under grant DE-FG02-95ER40899.

\end{document}